\documentclass[conference]{IEEEtran}  

\usepackage[draft]{graphicx}
\usepackage{graphicx}
\usepackage{amsmath}
\usepackage{amssymb}
\usepackage{booktabs}
\usepackage{enumerate}
\usepackage{algorithm}
\usepackage{algorithmic}
\usepackage{amsmath}
\usepackage{graphics}
\usepackage{epsfig}
\usepackage{subfigure}
\usepackage{mmap} 
\usepackage{multirow}

\usepackage{bbding}
\usepackage{amsmath}
\usepackage{amssymb}
\usepackage{ntheorem}
\usepackage{graphicx}
\usepackage{xcolor}

\psfull
\begin{document}
\title{Energy Harvesting Fairness in AN-aided \\ Secure MU-MIMO SWIPT Systems \\with Cooperative Jammer}

%
\author{\IEEEauthorblockN{Zhengyu Zhu$^{\dag}$, Zheng Chu$^\S$, Ning Wang$^{\dag}$$^\star$, 
Zhongyong Wang$^\dag$,  Inkyu Lee$^ \ddag$\\}
\IEEEauthorblockN{$^\dag$ School of Information Engineering, Zhengzhou University, China \\}
\IEEEauthorblockN{$^\star$ Department of Electrical and Computer Engineering, McMaster University, Canada \\}
\IEEEauthorblockN{$^\S$ 5G Innovation Center, Institute of Communication Systems, University of Surrey, UK\\}
\IEEEauthorblockN{$^\ddag$ School of Electrical Engineering, Korea University, Korea}
Email: zhuzhengyu6@gmail.com, andrew.chuzheng7@gmail.com, iezywang@zzu.edu.cn,  inkyu@korea.ac.kr }
\maketitle
\begin{abstract}

In this paper, we study a multi-user multiple-input-multiple-output secrecy simultaneous wireless information and power transfer (SWIPT) channel which consists of one transmitter, one cooperative jammer (CJ), multiple energy receivers (potential eavesdroppers, ERs), and multiple co-located receivers (CRs). We exploit the dual of artificial noise (AN) generation for facilitating efficient wireless energy transfer and secure transmission. Our aim is to maximize the minimum harvested energy among ERs and CRs subject to secrecy rate constraints for each CR and total transmit power constraint.
By incorporating norm-bounded channel uncertainty model, we propose a iterative algorithm based on sequential parametric convex approximation to find a near-optimal solution.
Finally, simulation results are presented to validate the performance of the proposed algorithm outperforms that of the conventional AN-aided scheme and CJ-aided scheme.

\end{abstract}

\IEEEpeerreviewmaketitle


\IEEEpeerreviewmaketitle
\setlength{\baselineskip}{1\baselineskip}
\newtheorem{definition}{Definition}
\newtheorem{fact}{Fact}
\newtheorem{assumption}{Assumption}
\newtheorem{theorem}{Theorem}
\newtheorem{lemma}{Lemma}
\newtheorem{corollary}{Corollary}
\newtheorem{proposition}{Proposition}
\newtheorem{example}{Example}
\newtheorem{remark}{Remark}

\section{Introduction}

Recently, wireless power transfer (WPT) has been a promising paradigm to scavenge energy from the radio frequency (RF) signals \cite{6 Grover10_ISIT_Shannon meets tesla WIPT}.
As a key technology for really perpetual communications, simultaneous wireless information and power transfer (SWIPT) has been an promising of interests in RF-enabled signal to provide power supplies for wireless networks, which have been studied in various scenarios \cite{7 Zhang13_MIMOSWIPT}-\cite{JCR_zhu_SWIPT}.
On the other hand, secrecy transmission,  especially physical-layer security (PLS), has extracted more and more attentions in 5G wireless networks \cite{zhujun_massive_MIMO_Secure_1}.

Specifically, PLS has been recognized as a important issue for SWIPT system due to its inherent characteristics make the wireless information more vulnerable to eavesdropping \cite{chenxiaoming_Secrecy_SWIPT}-\cite{17 Ng_robust_secureSWIPT}.
Moreover, the secure transmission with SWIPT schemes has also been investigated in the multi-user multiple-input-multiple-output (MU-MIMO) broadcasting \cite{15 Shi_Secrecy_MISO-BC SWIPT}.

It is noted that the assumption that perfect channel state information (CSI) is available at the transmitter in \cite{7 Zhang13_MIMOSWIPT}-\cite{9 HoonLee15_SWIPT_IC} \cite{chenxiaoming_Secrecy_SWIPT}-\cite{15 Shi_Secrecy_MISO-BC SWIPT}.  
In practice, it is not always possible to obtain perfect CSI at the transmitter due to the channel errors.  Secure communication with SWIPT would be more challenging with imperfect CSI at the transmitter.  Some robust optimization techniques have been constructed to secrecy SWIPT transmission under imperfect channel realization in \cite{Complexity_zhu}\cite{17 Ng_robust_secureSWIPT} \cite{19 Chu_SWIPT_MISO secrecy channel}-\cite{23 Qin_Cooperative_jamming aided robust secure}.
Considering the SWIPT scheme, an optimal transmit covariance matrix robust design has been proposed for MIMO secure channels with multi-antenna eavesdroppers \cite{17 Ng_robust_secureSWIPT}.

In addition, some of state-of-art techniques have been developed to introduce more interference to the eavesdroppers \cite{14Negi_08TWC_tichu_AN}-\cite{25 LiQ_13TSP_Spatially selective_AN}. Artificial noise (AN) technique has been used to embed the transmit beamforming to confuse the eavesdropper \cite{14Negi_08TWC_tichu_AN}. In secrecy SWIPT systems,  AN plays both the roles of an energy-carrying signal for WPT and protecting the secrecy information transmission, which has been considered as interfering the eavesdropper and harvesting power simultaneously in \cite{19 Chu_SWIPT_MISO secrecy channel}-\cite{TIFS_zhu}. In addition, to further increase the secrecy rates, jamming node has been introduced in the secrecy networks, which has the capability to improve the legitimate user's performance and prevent the eavesdroppers from intercepting the intended messages \cite{5 Lee_WPCN}-\cite{22 chuTVT}.
Based on the worst-case scheme, cooperative jamming signal was generated by a external cooperative jammer (CJ) node to interfere the eavesdropper and improve the secrecy rate in multiple-input-single-output (MISO) secure SWIPT system with wiretap channels\cite{23 Qin_Cooperative_jamming aided robust secure}.

When information receivers (IRs) and energy-harvesting receivers (ERs) are placed in a same cell, the ERs are normally assumed to be closer to the transmitter compared with IRs. This gives rise to a new information security issue in the SWIPT systems. In such a situation,
ERs have a possibility of eavesdropping the information sent to the IRs, and thus can become potential eavesdroppers \cite{13 Liuliang_Secrecy_SWIPT}
\cite{20 Khandaker_Masked beamforming} \cite{TIFS_zhu}.

In this paper, considering SWIPT, we investigate a secure transmission design problem in MU-MIMO secrecy system
with one multi-antenna transmitter, one multi-antenna CJ, multiple single-antenna co-located receiver (CR), and multiple multi-antenna ER. The CR employs power splitting (PS) scheme to split the received signals into two streams for ID and energy harvesting (EH) simultaneously.
Unlike \cite{24 Hongxing_jam_TSP} \cite{23 Qin_Cooperative_jamming aided robust secure}, our objective is to maximize the minimum of harvested energy (max-min HE) of both ERs and CRs subject to secrecy rate constraints for each CR and total transmit power constraint. Assuming the imperfect CSI case, we seek to the optimal transmission strategy to jointly optimize the AN-aided beamforming, the AN vector, the CJ vector and the PS ratio design.

Considering the worst-case scheme by incorporating the norm-bounded channel uncertainty model, we derive equivalent forms of secrecy rate constraint and the minimum of harvested energy. An iterative algorithm based on sequential parametric convex approximation (SPCA) is also addressed to recover a high quality rank-one beamforming solution to the original problem. Finally, the performance analysis are provided to verify that the proposed algorithm outperforms the  conventional scheme.

\textbf{\emph{Notation:}}
 $\otimes$ defines the Kronecker product.  $\mathbb{C}^{M\times L}$ and $\mathbb{H}^{M\times L}$ describe the space of ${M\times L}$ complex matrices and Hermitian matrices, respectively. $ \mathbb{H}_{+}$ equals the set of positive semi-definite Hermitian matrices, and  $ \mathbb{R}_{+}$ denotes the set of all nonnegative real numbers. For a matrix ${\mathbf{A}}$, ${\mathbf{A}}\succeq {\mathbf{0}}$ means
that ${\mathbf{A}}$ is positive semi-definite, and $\|\mathbf{{A}}\|_F$, $\rm{tr} ({\mathbf{A}})$, $|{\mathbf{A}}|$ and $\rm{rank}({\mathbf{A}})$ denote the Frobenius norm, trace, determinant, and the rank, respectively. ${{\emph{vec}}}(\mathbf{{A}})$ stacks the elements of $\mathbf{{A}}$ in a column vector. ${{{\mathbf{0}}}}_{M\times L}$ is a null matrix with ${M\times L}$ size.   $\mathrm{E}\{\cdot\}$ describes  expectation, and $\Re\{\cdot\}$ stands for the real part of a complex number. $[x]^{+}$ represents $\max\{x,0\}$ and $\lambda_{max}(\mathbf{{A}})$ indicates the maximum eigenvalue of $\mathbf{{A}}$.

\section{System Model and Problem Formulation} 
In this section, we consider a MU-MIMO secrecy channel, which consists of one multi-antenna transmitter, one multi-antenna CJ, $ K $ single-antenna CRs, and $ L $ multi-antenna ERs, where the CR employs the PS scheme to decode information and exploit power simultaneously. It is assumed that the transmitter and the CJ are equipped with $ N_{T} $ and $ N_{J} $ transmit antennas, and each ER has $ N_{E} $ receive antennas.
We denote $ {\mathbf{h}}_k \in \mathbb{C}^{N_{T}} $ as the channel vector between the transmitter and the $ k $-th CR, $ {\mathbf{H}}_l \in \mathbb{C}^{N_{T} \times N_{E}} $ as the channel matrix between the transmitter and the $ l $-th ER, $ {\mathbf{g}}_k \in \mathbb{C}^{N_{J}} $ as the channel vector between the CJ and the $ k $-th CR,  $ {\mathbf{G}}_l \in \mathbb{C}^{N_{J} \times N_{E}} $ as the channel matrix between the CJ and the $ l $-th ER, respectively. 

In order to improve the reliable transmission, the transmitter employs the transmit beamforming with AN, which acts as interference to the ERs and simultaneously provides energy to the CRs and ERs. Thus, the transmitter sends the confidential message ${\mathbf{x}}_k$ by using transmit beamforming with AN to the $k$-th CR  as
\begin{eqnarray}
{\mathbf{x}}_k = {{\mathbf{w}_k}}s_k + \textbf{\emph{z}},
\end{eqnarray}
where $ {{\mathbf{w}_k}} \in \mathbb{C}^{N_{T}}$ denotes the linear beamforming vector for the $k$-th CR at the transmitter, $ s_k  \in \mathcal{C} $ represents the information-bearing signal intended for the $k$-th CR satisfying $\mathbb{E}\{|s_k|^2\}=1$, and $ \textbf{\emph{z}} \in \mathbb{C}^{N_{T}}$ is the energy-carrying AN vector, which can also be composed by multiple energy beams.
The received signal at the $ k $-th CR and the $ l $-th ER can be expressed as
\begin{equation}\label{eq:1}
\begin{split}
&\!\! y_k =  {\mathbf{h}}_k^{H}\sum\nolimits_{k=1}^K {\mathbf{w}}_{k} s_{k} \!+\! {\mathbf{h}}_k^{H}{\mathbf{z}}  \!+\! {\mathbf{g}}_k^{H}{\mathbf{q}} s_{J} \!+\! n_k, ~ k = 1,...,K, \\
&\!\! \textbf{\emph{y}}_l  \!=\! {\mathbf{H}}_l^{H} \sum\nolimits_{k=1}^{K}{\mathbf{w}}_{k} s_{k} \!+\! {\mathbf{H}}_l^{H} {\mathbf{z}}  \!+\! {\mathbf{G}}_l^{H}{\mathbf{q}} s_{J} \!+\! {\mathbf{n}}_l, ~ l = 1,...,L,
\end{split}
\end{equation}
where $s_{J}$ is the cooperative jamming signal introduced by the CJ  satisfying $\mathbb{E}\{|s_J|^2\}=1$, $ {\mathbf{q}} \in \mathbb{C}^{N_{J}}$ indicates the CJ vector, $ n_k \sim \mathcal{CN}(0, \sigma_k^{2}) $ and $ {\mathbf{n}}_l \sim \mathcal{CN}(0, \sigma_l^{2}{{{\mathbf{I}}}}) $ stand for the additive Gaussian noise by the receive antenna at the $ k $-th CR and the $ l $-th ER.
In addition, each CR considers the PS scheme to manage the processes of ID and EH simultaneously. Based on this reason, the received signal at the $ k $-th CR is divided into ID and EH by PS ratio $ \rho_k\in (0, 1] $. Thus, the received signal for ID at the $ k $-th CR  can be given as
\begin{equation}\label{eq:2}
\begin{split}
y_k^{ID} & =  \sqrt{\rho_k}y_k \!+\! n_{p,k} \\
 &=  \sqrt{\rho_k} \big({\mathbf{h}}_k^{H}\sum\nolimits_{k = 1}^K {\mathbf{w}}_{k} s_{k}
 +  {\mathbf{h}}_k^{H}{\mathbf{z}} \!+\! {\mathbf{g}}_k^{H}{\mathbf{q}} s_{J} \!+\! n_k\big) \!+\! n_{p,k}, \forall k,
\end{split}
\end{equation}
where $ n_{p,k} \sim \mathcal{CN}(0, \delta_k^2) $ is the antenna noise introduced by signal process of the ID at the $k$-th CR \cite{7 Zhang13_MIMOSWIPT}.

We denote ${{\mathbf{W}}}_{k} = {\mathbf{w}}_{k}{\mathbf{w}}_{k}^{H}$ as the transmit covariance matrix,  ${{\mathbf{Z}}} = {\mathbf{z}}{\mathbf{z}}^{H}$ as the AN covariance matrix, and ${{\mathbf{Q}}} = {\mathbf{q}}{\mathbf{q}}^{H}$ as the CJ covariance matrix.
Hence, the achieved secrecy rate can be calculated by
\begin{equation}\label{eq:3}
\hat{R}_k  = \bigg[\log_2 \big( 1 + {\textrm{SINR}}_k\big) -\max_l~  C_{l,k}
 \bigg]^{+}, ~\forall k,
\end{equation}
where
\begin{equation*}\label{eq:rate_define}
\begin{split}
{\textrm{SINR}}_k & \!=\! \frac{\rho_k {\mathbf{h}}_k^{H}{{\mathbf{W}}}_{k}{\mathbf{h}}_k}{\rho_k (\sigma_k^{2}+  {\mathbf{h}}_k^{H}(\sum\nolimits_{j \neq k}{{\mathbf{W}}}_{j}+{{\mathbf{Z}}}){\mathbf{h}}_k + {\mathbf{g}}_k^{H}{{\mathbf{Q}}}{\mathbf{g}}_k) + \delta_k^2}, \\
C_{l,k} &\!=\! \log_2 \big| {{{\mathbf{I}}}} \!+\! ({\mathbf{H}}_l^{H}{{\mathbf{Z}}}{\mathbf{H}}_l \!+\! {\mathbf{G}}_l^{H}{{\mathbf{Q}}}{\mathbf{G}}_l \!+\! \sigma_l^{2}{{{\mathbf{I}}}})^{-1} {\mathbf{H}}_l^{H}{{\mathbf{W}}}_{k}{\mathbf{H}}_l \big|.
\end{split}
\end{equation*}
Thus, the harvested power at the $k$-th CR and the $ l $-th ER are written as
\begin{equation}\label{eq:4}
\begin{split}
\!\!\!\! E_{c,k} & \!=\!  \eta_{c,k}(1 \!-\! \rho_k)\big({\mathbf{h}}_k^{H}\big(\sum\nolimits_{j=1}^{K}{{\mathbf{W}}}_{k} \!+\! {{\mathbf{Z}}}\big){\mathbf{h}}_k \!+\! {\mathbf{g}}_k^{H}{{\mathbf{Q}}}{\mathbf{g}}_k \!+\! \sigma_k^{2}\big), \\
\!\!\!\! E_{e,l} & \!=\!  \eta_{e,l}\big(\textrm{tr}({\mathbf{H}}_l^{H}(\sum\nolimits_{k=1}^{K} {{\mathbf{W}}}_{k}\!+\!{{\mathbf{Z}}}){\mathbf{H}}_l)
\!+\!  \textrm{tr}({\mathbf{G}}_l^{H}{{\mathbf{Q}}}{\mathbf{G}}_l) \!+\! N_{E}\sigma_l^{2}\big),
\end{split}
\end{equation}
where $\eta_{c,k}$ and $\eta_{e,l}$ denote the energy conversion efficiency of the $k$-th CR and the $ l $-th ER.

Due to channel estimation and quantization errors, it may not be possible to achieve the perfect CSI at the transmitter in
practice. In this section, our aim is to jointly optimize the max-min worst-case HE formulation at the imperfect CSI case.
Now, we adopt the imperfect CSI case under the norm-bounded channel uncertainty model \cite{17 Ng_robust_secureSWIPT}\cite{19 Chu_SWIPT_MISO secrecy channel}\cite{20 Khandaker_Masked beamforming}. Specifically,
the actual channels $ {\mathbf{h}}_k $, ${\mathbf{H}}_l$,  $ {\mathbf{g}}_k $, and  ${\mathbf{G}}_l$  can be given as
\begin{equation}\label{eq:5_1}
 \begin{split}
{\mathbf{h}}_k & =  {\bar{\mathbf{h}}}_k \!+\! {{\mathbf{e}}}_k, \forall k,  ~~{\mathbf{H}}_l  =  {\bar{\mathbf{H}}}_l \!+\! {{\mathbf{E}}}_l, \forall l, \\
{\mathbf{g}}_k & = {\bar{\mathbf{g}}}_k \!+\!  {\tilde{{\mathbf{e}}}}_k,\forall k,~~ {\mathbf{G}}_l  = {\bar{\mathbf{G}}}_l \!+\! {\tilde{{\mathbf{E}}}}_l, \forall l,
\end{split}
\end{equation}
where ${\bar{\mathbf{h}}}_k, {\bar{\mathbf{g}}}_k, {\bar{\mathbf{H}}}_l$, and ${\bar{\mathbf{G}}}_l$ denote the estimated channel available at the transmitter and the CJ, respectively, and $ {{\mathbf{e}}}_k, {\tilde{{\mathbf{e}}}}_k, {{\mathbf{E}}}_l$, and $ {\tilde{{\mathbf{E}}}}_l$ are the channel errors, bounded as $\|{{\mathbf{e}}}_k\|_{2} \leq \varepsilon_k$, $\|{\tilde{{\mathbf{e}}}}_k\|_{2} \leq \tilde{\varepsilon}_k$, $\|{{\mathbf{E}}}_l\|_{F} \leq \theta_l $, and $ \|{\tilde{{\mathbf{E}}}}_l\|_{F} \leq \tilde{\theta}_l$, respectively.

In this paper, our aim is to maximize the minimum of the total harvested power among the all CRs and ERs subject to the secrecy rate constraint and the total transmit power constraint at the transmitter and the CJ power constraint. By taking the above channel model into account, the transmit beamforming design is formulated as a multi-object optimization problem which can be given by
\begin{subequations}\label{eq:6}
\begin{eqnarray}
&&\!\!\! \max_{\rho_k,{\kern 1pt}\{{{\mathbf{W}}}_{k}\},{\kern 1pt}{{\mathbf{Z}}},{\kern 1pt}{{\mathbf{Q}}} } ~~~~  \hat{E}_{c,k}  + \hat{E}_{e,l}  \label{eq:6a} \\
&& \!\!\!\!\!\!\!\!\!\!\!\!\!\!\!\!\!\!\!\! s.t. \mathop {\min }_{\begin{array}{*{20}{c}}
\|{{\mathbf{e}}}_k\| \leq \varepsilon_k\\
\|{\tilde{{\mathbf{e}}}}_k\| \leq \tilde{\varepsilon}_k
\end{array}}  C_k \!-\! \mathop {\max }\limits_{\begin{array}{*{20}{c}}
{\|{{\bf{E}}_l}\|{_F} \leq {\theta_l}}\\
{\|{{\tilde {\bf{E}}}_l}\|{_F} \leq {\tilde{\theta}_l}}
\end{array}}  C_l \geq \bar{R}_{s}, \forall k, \label{eq:6b}\\
&&\!\!\!\!\!\!\!\!\!\!\!\!\!\!\!\!\!\! ~\sum\nolimits_{k=1}^{K}\textrm{tr}({{\mathbf{W}}}_{k}) \!+\! \textrm{tr}({{\mathbf{Z}}}) \leq P_{T},~\textrm{tr}({{\mathbf{Q}}}) \leq P_{J}, \label{eq:6c}\\ 
&&\!\!\!\!\!\!\!\!\!\!\!\!\!\!\!\!\!\!\!\!\!\!\! 1 \geq\rho_k >0, ~ {{\mathbf{W}}}_{k} \succeq {{{\mathbf{0}}}}, ~{{\mathbf{Z}}} \succeq {{{\mathbf{0}}}},~{{\mathbf{Q}}} \succeq {{{\mathbf{0}}}},~\textrm{rank}({{\mathbf{W}}}_{k}) = 1, \label{eq:6d}
\end{eqnarray}
\end{subequations}
where
\begin{equation*}
 \begin{split}
&\hat{E}_{c,k} \triangleq \mathop {\min }\limits_{
{\|{{\mathbf{e}}_k}\|  \leq {\varepsilon _k}},
{\|{{ \tilde{{\mathbf{e}}}}_k}\|  \leq {{\tilde \varepsilon }_k}}} \min_{c,k} ~\tau E_k, ~\forall k,\\
&\hat{E}_{e,l} \triangleq \mathop {\min }\limits_{
{\|{{\bf{E}}_{e,l}}\|{_F} \leq {\theta_l}},
{\|{{\widetilde {\bf{E}}}_l\|}{_F} \leq {\tilde{\theta}_l}}}  \min_{l}~ (1-\tau) E_{e,l}, ~\forall l,
 \end{split}
\end{equation*}
$\tau$ is the priority parameter, $\bar{R}_{s}$ stands for a given secrecy rate threshold, and $P_{T}$ and $P_{J}$ denote the available power budget at the transmitter and the CJ, respectively.  
Variable  $\tau \geq 0$  reflects the preference of the system operator. Problem \eqref{eq:6} is non-convex due to the secrecy rate constraint and the objective function, and thus cannot be solved directly.

\section{Proposed Robust Design Method}
In order to circumvent the roust max-min HE problem, we transform  problem  \eqref{eq:6} by introducing two slack variables $\bar{E}_{s}$ and $\bar{E}_{e}$   into
\begin{subequations}\label{eq:7}
\begin{eqnarray}
&&\!\!\!\!\!\! \max_{\rho_k,{\kern 1pt}\{{{\mathbf{W}}}_{k}\},{\kern 1pt}{{\mathbf{Z}}},{\kern 1pt}{{\mathbf{Q}}}, {\kern 1pt}\bar{E}_{s},{\kern 1pt}\bar{E}_{e} } ~~  \tau \bar{E}_{s}  +   (1-\tau) \bar{E}_{e} \label{eq:7a} \\
&&\!\!\!\!\!\!s.t.~~ \min_{\|{{\mathbf{e}}}_k\| \leq \varepsilon_k,\|{\tilde{{\mathbf{e}}}}_k\| \leq \tilde{\varepsilon}_k} \min_{k}~ E_{c,k} \geq \bar{E}_{s}, ~ \forall k, \label{eq:7b} \\
&& ~~\min_{\|{{\mathbf{E}}}_l\|_{F} \leq \theta_l,\|{\tilde{{\mathbf{E}}}}_l\|_{F} \leq \tilde{\theta}_l}\min_{l} ~ E_{e,l} \geq \bar{E}_{e},~ \forall l, \label{eq:7c} \\
&&\! ~~~ \eqref{eq:6b}, ~~\eqref{eq:6c}, ~~\eqref{eq:6d}.\nonumber
\end{eqnarray}
\end{subequations}
Problem \eqref{eq:7} is still non-convex in terms of \eqref{eq:7b}, \eqref{eq:7c} and \eqref{eq:6b}.

Now,  let us consider another formulation of  problem \eqref{eq:6}
based on SPCA method \cite{SPCA_firstorder}.
The optimization framework can also be recast as a convex form by incorporating
channel uncertainties.
First, by applying the matrix inequality $| \textbf{I} + \textbf{A}| \geq 1+ \rm{tr}(\textbf{A})$ \cite{Matrix Cookbook}, the robust secrecy rate constraint \eqref{eq:6b} can be relaxed as
\begin{equation}\label{eq:zzy_1}
 C_k - \log_2 \bigg(1 \!+\! \frac{{\rm{tr}}\left({\mathbf{H}}_l^{H}{{\mathbf{W}}}_{k}{\mathbf{H}}_l\right)}
 {{\sigma_l^2 \!+\! {\rm{tr}}\left({\mathbf{H}}_l^{H}{{\mathbf{Z}}}{\mathbf{H}}_l \!+\! {\mathbf{G}}_l^{H}{{\mathbf{Q}}}{\mathbf{G}}_l\right)}}\bigg) \!\geq\!  \bar{R}_{s}.
\end{equation}

To make the constraint \eqref{eq:zzy_1} tractable, we introduce two slack variables $r_1 > 0$ and $r_2 > 0$,
the robust secrecy rate  \eqref{eq:zzy_1} can be equivalently relaxed as
\begin{subequations}\label{eq:99}
\begin{eqnarray}
&&\!\!\!\!\!\!\!\!\!\!\!\!\!\!\!\!\!\!\!\!\!\!\!\!  \log( r_1r_2 ) \geq  \bar{R}_{s},  \label{eq:99a}\\
&&\!\!\!\!\!\!\!\!\!\!\!\!\!\!\!\!\!\!\!\!\!\!\!\!\!  1 \!+\! \frac{{\rho_k} {\mathbf{h}}_k^{H}{{\mathbf{W}}}_{k}{\mathbf{h}}_k}{{\rho_k} (\ \sigma_k^{2}\!+\!  {\mathbf{h}}_k^{H}(\sum\nolimits_{j \neq k}{{\mathbf{W}}}_{j}\!+\!{{\mathbf{Z}}}){\mathbf{h}}_k + {\mathbf{g}}_k^{H}{{\mathbf{Q}}}{\mathbf{g}}_k) \!+\! {\delta_k^2}}  \!\geq\! r_1, \label{eq:99b}\\
&&\!\!\!\!\!\!\!\!\!\!\!\!\!\!\!\!\!\!\!\!\!\!\!\!
 1 + \frac{{\rm{tr}}\left({\mathbf{H}}_l^{H}{{\mathbf{W}}}_{k}{\mathbf{H}}_l\right)}
 {{\sigma_l^2 + {\rm{tr}}\left({\mathbf{H}}_l^{H}{{\mathbf{Z}}}{\mathbf{H}}_l \!+\! {\mathbf{G}}_l^{H}{{\mathbf{Q}}}{\mathbf{G}}_l\right)}}  \leq \frac{1}{r_2}, \forall l. \label{eq:99c}
\end{eqnarray}
\end{subequations}
Then, we can be further simplify \eqref{eq:99} as
\begin{subequations} \label{eq:10}
\begin{eqnarray}
&&\!\!\!\!\!\!\!\!\!\!\!\!\!\!\!\!\!\!\!\!\!\!\!\!   r_1r_2  \geq  2^{\bar{R}_{s}},  \label{eq:10a}\\
&&\!\!\!\!\!\!\!\!\!\!\!\!\!\!\!\!\!\!\!\!\!\!\!\!\!\!\!\!   \frac{ {\mathbf{h}}_k^{H}{{\mathbf{W}}}_{k}{\mathbf{h}}_k}{\ \sigma_k^{2}+  {\mathbf{h}}_k^{H}(\sum\nolimits_{j \neq k}{{\mathbf{W}}}_{j}\!+\!{{\mathbf{Z}}}){\mathbf{h}}_k +  {\mathbf{g}}_k^{H}{{\mathbf{Q}}}{\mathbf{g}}_k + \frac{{\delta_k^2}}{{\rho_k}}}  \geq {r_1} \!-\! 1, \forall l,\label{eq:10b}\\
&&\!\!\!\!\!\!\!\!\!\!\!\!\!\!\!\!\!\!\!\!\!\!\!\!
  \frac{{\sigma_l^2 + {\rm{tr}}\left({\mathbf{H}}_l^{H}{{\mathbf{Z}}}{\mathbf{H}}_l \!+\! {\mathbf{G}}_l^{H}{{\mathbf{Q}}}{\mathbf{G}}_l\right)}}
 {{\sigma_l^2 + {\rm{tr}}\left({\mathbf{H}}_l^{H}({{\mathbf{Z}}}+{{\mathbf{W}}}_{k}){\mathbf{H}}_l \!+\! {\mathbf{G}}_l^{H}{{\mathbf{Q}}}{\mathbf{G}}_l\right)}}  \geq  {{r_2}}, \forall k. \label{eq:10c}
\end{eqnarray}
\end{subequations}

The inequality constraint \eqref{eq:10a} is equivalent to ${2^{\bar{R}_{s}+2}} + (r_1-r_2)^2 \leq (r_1+r_2)^2$,
which can be converted into a conic quadratic-representable function form as
\begin{equation} \label{eq:11a}
\left\|\left[\sqrt{2^{\bar{R}_{s}+2}}~~~~ r_1-r_2 \right]\right\|\leq r_1+r_2.
\end{equation}
Design ${\bar{\mathbf{H}}}_k \triangleq {\bar{\mathbf{h}}}_k^{H}{\bar{\mathbf{h}}}_k$,  ${\bar{\mathbf{G}}}_k \triangleq  {\bar{\mathbf{g}}}_k^{H}{\bar{\mathbf{g}}}_k$, ${\hat{\mathbf{H}}}_l  \triangleq  {\bar{\mathbf{H}}}_l^{H}{\bar{\mathbf{H}}}_l$, and ${\hat{\mathbf{G}}}_l  \triangleq  {\bar{\mathbf{G}}}_l^{H}{\bar{\mathbf{G}}}_l$.
The inequalities in \eqref{eq:10b} and \eqref{eq:10c} can be rearranged, respectively, which give
\begin{subequations}
\begin{eqnarray}
&&\!\!\!\!\!\!\!\!\!\!\!\!\!\! \ \sigma_k^{2}\!+\! \sum\nolimits_{j \neq k}{\mathbf{w}}_{j}^H({\bar{\mathbf{H}}}_k\!+\!\Delta_k){\mathbf{w}}_{j} \!+\! {\mathbf{z}}^{H}({\bar{\mathbf{H}}}_k\!+\!\Delta_k){\mathbf{z}} \!+\! {\mathbf{q}}^{H}({\bar{\mathbf{G}}}_k\!+\!\Phi_k){\mathbf{q}} \nonumber \\
&&\!\!\!\!\!\! \!+\! \frac{{\delta_k^2}}{{\rho_k}}  \leq \frac{{\mathbf{w}}_{k}^H({\bar{\mathbf{H}}}_k\!+\!\Delta_k){\mathbf{w}}_{k}}{{r_1} -1}, \forall k, \label{eq:25b}\\
&&\!\!\! {{\sigma_l^2 \!+\! {\mathbf{z}}^{H}({\hat{\mathbf{H}}}_l\!+\!\Upsilon_l  ){\mathbf{z}} \!+\! {\mathbf{w}}_k^{H}({\hat{\mathbf{H}}}_l\!+\!\Upsilon_l  ){\mathbf{w}}_k \!+\! {\mathbf{q}}^{H}({\hat{\mathbf{G}}}_l\!+\! \Psi_l  ){\mathbf{q}}}} \nonumber\\
&&\!\!\! \leq \frac{{\sigma_l^2 \!+\! {\mathbf{z}}^{H}({\hat{\mathbf{H}}}_l\!+\!\Upsilon_l  ){\mathbf{z}} \!+\! {\mathbf{q}}^{H}({\hat{\mathbf{G}}}_l\!+\! \Psi_l  ){\mathbf{q}}}}{r_2}, \forall l, \label{eq:25c}
\end{eqnarray}
\end{subequations}
where $\Delta_k ={\bar{\mathbf{h}}}_k{{\mathbf{e}}}_k^H+{{\mathbf{e}}}_k{\bar{\mathbf{h}}}_k^H+{{\mathbf{e}}}_k{{\mathbf{e}}}_k^H$, $\Phi_k = {\bar{\mathbf{g}}}_k{\tilde{{\mathbf{e}}}}_k^H+{\tilde{{\mathbf{e}}}}_k{\bar{\mathbf{g}}}_k^H+{\tilde{{\mathbf{e}}}}_k{\tilde{{\mathbf{e}}}}_k^H$, $\Upsilon_l = {\bar{\mathbf{H}}}_l {{\mathbf{E}}}_l^H + {{\mathbf{E}}}_l{\bar{\mathbf{H}}}_l^H +{{\mathbf{E}}}_l{{\mathbf{E}}}_l^H$, and $\Psi_l  = {\bar{\mathbf{G}}}_l  {\tilde{{\mathbf{E}}}}_l^H + {\tilde{{\mathbf{E}}}}_l{\bar{\mathbf{G}}}_l^H + {\tilde{{\mathbf{E}}}}_l {\tilde{{\mathbf{E}}}}_l^H$.
which stand for
the CSI uncertainty.
It is straightforward to show that
\begin{equation}
\begin{split}
\displaystyle  \left\| \Delta_k \right\|_F & \le \| {\bar{\mathbf{h}}}_k{{\mathbf{e}}}_k^H\|_F + \|{{\mathbf{e}}}_k{\bar{\mathbf{h}}}_k^H\|_F + \| {{\mathbf{e}}}_k{{\mathbf{e}}}_k^H \|_F\\
 \displaystyle & \le  {\| {\bar{\mathbf{h}}}_k \|}   \| {{\mathbf{e}}}_k^H  \| + \|{{\mathbf{e}}}_k\| \| {\bar{\mathbf{h}}}_k^H \|{\rm{ + }}{\| {{\mathbf{e}}}_k \|^2}\\
 \displaystyle  &  =  \varepsilon_k^2 + 2\varepsilon_k\|{\bar{\mathbf{h}}}_k\|,
\end{split}
\end{equation}
\begin{equation}
\begin{split}
\displaystyle  \left\| \Phi_k \right\|_F & \le \| {\bar{\mathbf{g}}}_k{\tilde{{\mathbf{e}}}}_k^H\|_F + \| {\tilde{{\mathbf{e}}}}_k{\bar{\mathbf{g}}}_k^H\|_F + \| {\tilde{{\mathbf{e}}}}_k{\tilde{{\mathbf{e}}}}_k^H \|_F\\
 \displaystyle & \le  {\| {\bar{\mathbf{g}}}_k \|}   \| {\tilde{{\mathbf{e}}}}_k^H  \| + \|{\tilde{{\mathbf{e}}}}_k\| \| {\bar{\mathbf{g}}}_k^H \|{\rm{ + }}{\| {\tilde{{\mathbf{e}}}}_k \|^2}\\
 \displaystyle  &  =  \tilde{\varepsilon}_k^2 + 2\tilde{\varepsilon}_k\|{\bar{\mathbf{g}}}_k\|,
\end{split}
\end{equation}
\begin{equation}
\begin{split}
\displaystyle  \left\| {{\Upsilon_l}} \right\|_F  & \le \| {\bar{\mathbf{H}}}_l {{\mathbf{E}}}_l^H \|_F{\rm{ + }}\| {{\mathbf{E}}}_l{\bar{\mathbf{H}}}_l ^H \|_F{\rm{ + }}\| {{\mathbf{E}}}_l{{\mathbf{E}}}_l^H \|_F\\
 \displaystyle & \le  {\| {\bar{\mathbf{H}}}_l  \|_F}   \| {{\mathbf{E}}}_l^H  \|_F + {{\mathbf{E}}}_l\| _F\| {\bar{\mathbf{H}}}_l ^H \|_F{\rm{ + }}{\| {{\mathbf{E}}}_l \|_F^2}\\
 \displaystyle  &  =  \theta_l^2 + 2\theta_l\|{\bar{\mathbf{H}}}_l \|_F,
\end{split}
\end{equation}
\begin{equation}
\begin{split}
\displaystyle  \left\| {{\Psi_l}} \right\|_F  & \le \| {\bar{\mathbf{G}}}_l {\tilde{{\mathbf{E}}}}_l^H \|_F{\rm{ + }}\| {\tilde{{\mathbf{E}}}}_l{\bar{\mathbf{G}}}_l ^H \|_F{\rm{ + }}\| {\tilde{{\mathbf{E}}}}_l{\tilde{{\mathbf{E}}}}_l^H \|_F\\
 \displaystyle & \le  {\| {\bar{\mathbf{G}}}_l  \|_F}   \| {\tilde{{\mathbf{E}}}}_l^H  \|_F + \|{\tilde{{\mathbf{E}}}}_l\| _F\| {\bar{\mathbf{G}}}_l ^H \|_F{\rm{ + }}{\| {\tilde{{\mathbf{E}}}}_l \|_F^2}\\
 \displaystyle  &  =  \tilde{\theta}_l^2 + 2\tilde{\theta}_l\|{\bar{\mathbf{G}}}_l \|_F.
\end{split}
\end{equation}
Note that ${{\Delta _k}}$, $\Phi_k$, ${{\Upsilon_l}}$,  and ${{\Psi_l}}$ are  norm-bounded matrices as $\left\| {{\Delta _k}} \right\|_F \leq \xi_k$,  $\left\| {{\Phi _k}} \right\|_F \leq \tilde{\xi}_k$,  $\left\| {{\Upsilon_l}} \right\|_F \leq \alpha_l$, and $\left\| {{\Psi_l}} \right\|_F \leq \tilde{\alpha}_l$, where
$\xi_k = \varepsilon_k^2 + 2\varepsilon_k\|{\bar{\mathbf{h}}}_k\|$, $\tilde{\xi}_k = \tilde{\varepsilon}_k^2 + 2\tilde{\varepsilon}_k\|{\bar{\mathbf{g}}}_k\|$, $\alpha_l =\theta_l^2 + 2\theta_l\|{\bar{\mathbf{H}}}_l \|_F$, and $\tilde{\alpha}_l = \tilde{\theta}_l^2 + 2\tilde{\theta}_l\|{\bar{\mathbf{G}}}_l \|_F$.

According to \cite{loose_approxi}, we can minimize constraint \eqref{eq:10b} by maximizing the left-hand side (LHS) of \eqref{eq:25b} while minimizing its the  right-hand
side (RHS). Then \eqref{eq:25b} and \eqref{eq:25c} can be approximately rewritten as, respectively,
\begin{equation} \label{eq:zzy10}
\begin{split}
&\!\!\!\!\!\!\max_{\left\| {{\Delta _k}} \right\|_F \leq \xi_k,
\left\| {{\Phi _k}} \right\|_F \leq \tilde{\xi}_k} \ \sigma_k^{2}\!+\! \sum\nolimits_{j \neq k}{\mathbf{w}}_{j}^H({\bar{\mathbf{H}}}_k\!\!+\!\!\Delta_k){\mathbf{w}}_{j} \!+\! {\mathbf{z}}^{H}({\bar{\mathbf{H}}}_k\!+\!\Delta_k){\mathbf{z}}   \\
&\!\!\!\!\!\! \!+\! {\mathbf{q}}^{H}({\bar{\mathbf{G}}}_k\!+\!\Phi_k){\mathbf{q}}\!+\! \frac{{\delta_k^2}}{{\rho_k}} \leq \min_{\left\| {{\Delta _k}} \right\|_F \leq \xi_k} ~   \frac{{\mathbf{w}}_{k}^H({\bar{\mathbf{H}}}_k\!+\!\Delta_k){\mathbf{w}}_{k}}{{r_1} -1}, \forall k,
\end{split}
\end{equation}
\begin{equation} \label{eq:zzy11}
\begin{split}
& \max_{ \left\| {{\Upsilon_l}} \right\|_F \leq \alpha_l, \left\| {{\Psi_l}} \right\|_F \leq \tilde{\alpha}_l} ~\sigma_l^2 \!+\! {\mathbf{z}}^{H}({\hat{\mathbf{H}}}_l\!+\!\Upsilon_l  ){\mathbf{z}} \!+\! {\mathbf{w}}_k^{H}({\hat{\mathbf{H}}}_l+\Upsilon_l  ){\mathbf{w}}_k \\
& ~~~~~~~~~~~~~~~~~~~~~~~~  \!+\! {\mathbf{q}}^{H}({\hat{\mathbf{G}}}_l+ \Psi_l  ){\mathbf{q}}  \\
&\!\!\!\!\!\! \leq  \min_{ \left\| {{\Upsilon_l}} \right\|_F \leq \alpha_l, \left\| {{\Psi_l}} \right\|_F \leq \tilde{\alpha}_l} \frac{{\sigma_l^2 \!+\! {\mathbf{z}}^{H}({\hat{\mathbf{H}}}_l\!+\!\Upsilon_l  ){\mathbf{z}} \!+\! {\mathbf{q}}^{H}({\hat{\mathbf{G}}}_l\!+\! \Psi_l  ){\mathbf{q}}}}{r_2}, \forall l.
\end{split}
\end{equation}

In order to minimize the RHS of \eqref{eq:zzy10} and \eqref{eq:zzy11}, a loose approximation \cite{loose_approxi} is applied, which gives
\begin{equation} \label{eq:zzy12}
\begin{split}
&\!\!\!\!\!\!\!\!\! \min_{\left\| {{\Delta _k}} \right\|_F \leq \xi_k} ~   \frac{{\mathbf{w}}_{k}^H({\bar{\mathbf{H}}}_k\!+\!\Delta_k){\mathbf{w}}_{k}}{{r_1} -1} \geq \frac{{\mathbf{w}}_{k}^H({\bar{\mathbf{H}}}_k-\xi_k{\mathbf{I}}){\mathbf{w}}_{k}}{{r_1} -1}, \forall k,\\
&\!\!\!\!\!\!\!\!\! \min_{ \left\| {{\Upsilon_l}} \right\|_F \leq \alpha_l, \left\| {{\Psi_l}} \right\|_F \leq \tilde{\alpha}_l} \frac{{\sigma_l^2 \!+\! {\mathbf{z}}^{H}({\hat{\mathbf{H}}}_l\!+\! \Upsilon_l  ){\mathbf{z}} \!+\! {\mathbf{q}}^{H}({\hat{\mathbf{G}}}_l\!+\!  \Psi_l  ){\mathbf{q}}}}{r_2} \\
& ~~~~~~~~~~~~~~~\geq \frac{{\sigma_l^2 \!+\!  {\mathbf{z}}^{H}({\hat{\mathbf{H}}}_l\!-\! \alpha_l{\mathbf{I}}  ){\mathbf{z}} \!+\! {\mathbf{q}}^{H}({\hat{\mathbf{G}}}_l\!-\!  \tilde{\alpha}_l{\mathbf{I}}){\mathbf{q}}}}{r_2}.
\end{split}
\end{equation}

Using similar technique to the LHS of \eqref{eq:zzy10} and \eqref{eq:zzy11} yields
 \begin{equation}  \label{eq:zzy13}
 \begin{split}
&  \max_{\left\| {{\Delta _k}} \right\|_F \leq \xi_k,
\left\| {{\Phi _k}} \right\|_F \leq \tilde{\xi}_k} ~  \sigma_k^{2}\!+\!  \sum\nolimits_{j \neq k}{\mathbf{w}}_{j}^H({\bar{\mathbf{H}}}_k\!+\! \Delta_k){\mathbf{w}}_{j} \!+\!  {\mathbf{z}}^{H}({\bar{\mathbf{H}}}_k \\
&\!+\! \Delta_k){\mathbf{z}} \!+\!  {\mathbf{q}}^{H}({\bar{\mathbf{G}}}_k\!+\! \Phi_k){\mathbf{q}}\!+\!  \frac{{\delta_k^2}}{{\rho_k}} \leq  \sigma_k^{2}\!+\!  \sum\nolimits_{j \neq k}{\mathbf{w}}_{j}^H({\bar{\mathbf{H}}}_k\!+\! \xi_k{\mathbf{I}}){\mathbf{w}}_{j}  \\
&\!+\!  {\mathbf{z}}^{H}({\bar{\mathbf{H}}}_k\!+\! \xi_k{\mathbf{I}}){\mathbf{z}}\!+\!  {\mathbf{q}}^{H}({\bar{\mathbf{G}}}_k\!+\! \tilde{\xi}_k{\mathbf{I}}){\mathbf{q}}\!+\!  \frac{{\delta_k^2}}{{\rho_k}}, \forall k,
\end{split}
\end{equation}
\begin{equation}  \label{eq:zzy14}
\begin{split}
&\!  \max_{ \left\| {{\Upsilon_l}} \right\|_F \leq \alpha_l, \left\| {{\Psi_l}} \right\|_F \leq \tilde{\alpha}_l} ~\sigma_l^2 \!+\!  {\mathbf{z}}^{H}({\hat{\mathbf{H}}}_l\!+\! \Upsilon_l  ){\mathbf{z}} \!+\! {\mathbf{w}}_k^{H}({\hat{\mathbf{H}}}_l\!+\! \Upsilon_l  ){\mathbf{w}}_k  \\
&~~~~~~~~~~~~~~~~~\!+\! {\mathbf{q}}^{H}({\hat{\mathbf{G}}}_l\!+\!  \Psi_l  ){\mathbf{q}} \leq  ~\sigma_l^2 \!+\!  {\mathbf{z}}^{H}({\hat{\mathbf{H}}}_l\!+\! \alpha_l{\mathbf{I}}  ){\mathbf{z}} \\
&~~~~~~~~~~~~~~~~~\!+\! {\mathbf{w}}_k^{H}({\hat{\mathbf{H}}}_l\!+\! \alpha_l{\mathbf{I}}  ){\mathbf{w}}_k \!+\! {\mathbf{q}}^{H}({\hat{\mathbf{G}}}_l\!+\!  \tilde{\alpha}_l{\mathbf{I}}  ){\mathbf{q}}, \forall l.
\end{split}
\end{equation}


From \eqref{eq:zzy10}-\eqref{eq:zzy14}, \eqref{eq:25b} and \eqref{eq:25c} can be given as,  respectively,
\begin{equation}\label{eq:zzy15}
\begin{split}
&\! \sigma_k^{2}+ \sum\nolimits_{j \neq k}{\mathbf{w}}_{j}^H({\bar{\mathbf{H}}}_k\!+\! \xi_k{\mathbf{I}}){\mathbf{w}}_{j} \!+\!  {\mathbf{z}}^{H}({\bar{\mathbf{H}}}_k\!+\! \xi_k{\mathbf{I}}){\mathbf{z}}  \\
& \!+\!  {\mathbf{q}}^{H}({\bar{\mathbf{G}}}_k\!+\! \tilde{\xi}_k{\mathbf{I}}){\mathbf{q}} \!+\!  \frac{{\delta_k^2}}{{\rho_k}} \leq
\frac{{\mathbf{w}}_{k}^H{\bar{\mathbf{H}}}_{\xi_s,k}{\mathbf{w}}_{k}}{{r_1} -1}, \forall k,
\end{split}
\end{equation}
\begin{equation}\label{eq:zzy16}
\begin{split}
&\!\!\!\!\!\!{{\sigma_l^2 \!+\!  {\mathbf{z}}^{H}({\hat{\mathbf{H}}}_l\!+\! \alpha_l{\mathbf{I}}  ){\mathbf{z}} \!+\! {\mathbf{w}}_k^{H}({\hat{\mathbf{H}}}_l\!+\! \alpha_l{\mathbf{I}}  ){\mathbf{w}}_k \!+\! {\mathbf{q}}^{H}({\hat{\mathbf{G}}}_l\!+\!  \tilde{\alpha}_l{\mathbf{I}}  ){\mathbf{q}}}}\\
&\!\!\!\!\!\!\leq  \frac{{\sigma_l^2 \!+\!  {\mathbf{z}}^{H}{\hat{\mathbf{H}}}_{\xi_e,l}{\mathbf{z}} \!+\! {\mathbf{q}}^{H}{\hat{\mathbf{G}}}_{\tilde{\xi}_e,l}{\mathbf{q}}}}{r_2} , \forall k, l,
\end{split}
\end{equation}
where ${\bar{\mathbf{H}}}_{\xi_s,k} \!=\! {\bar{\mathbf{H}}}_k \!-\!\xi_k{\mathbf{I}}$, ${\hat{\mathbf{H}}}_{\xi_e,l} \!=\! {\hat{\mathbf{H}}}_l \!-\!\alpha_l{\mathbf{I}}$  and  ${\hat{\mathbf{G}}}_{\tilde{\xi}_e,l} \!=\! {\hat{\mathbf{G}}}_l \!-\! \tilde{\alpha}_l{\mathbf{I}}  $.
We observe that these two constraints are non-convex, but the RHS of both (\ref{eq:zzy15}) and (\ref{eq:zzy16}) have the function form of quadratic-over-linear, which are convex functions \cite{26 Boyd_convex}.
Based on the idea of the constrained convex procedure \cite{SPCA_firstorder},
these quadratic-over-linear functions can be replaced by their first-order expansions, which transforms the problem into convex programming.
Specifically, we define
\begin{eqnarray}\label{eq:zzy1}
f_{\mathbf{{A}},a}(\mathbf{w},t) = \frac{{{\mathbf{w}^{H}\mathbf{{A}}\mathbf{w}}}}{t-a},
\end{eqnarray}
where $\mathbf{{A}} \succeq {\mathbf{0}}$ and $t \geq a$. At a certain point $(\mathbf{\tilde{w}}, \tilde{t})$, the first-order Taylor expansion of \eqref{eq:zzy1}
 is given by
\begin{eqnarray}\label{eq:zzy2}
F_{\mathbf{{A}},a}(\mathbf{w},t,\mathbf{\tilde{w}},\tilde{t}) = \frac{2\Re{\{\mathbf{\tilde{w}}^{H}\mathbf{{A}}\mathbf{w}\}}}{\tilde{t}-a} - \frac{{{\mathbf{\tilde{w}}^{H}\mathbf{{A}}\mathbf{\tilde{w}}}}}{(\tilde{t}-a)^2}(t-a).
\end{eqnarray}

By using the above results of Taylor expansion, for the points $(\mathbf{\tilde{w}}_k, \tilde{r}_1)$, $(\mathbf{\tilde{z}}, \tilde{r}_2)$ and $( \mathbf{\tilde{q}}, \tilde{r}_2)$, we can transform \eqref{eq:zzy15} and \eqref{eq:zzy16} into convex forms,  respectively, as
\begin{subequations} \label{eq:zzy17}
\begin{eqnarray}
&&\!\!\!\!\!\!\!\!\!\!   \sigma_k^{2}+ \sum\nolimits_{j \neq k}{\mathbf{w}}_{j}^H({\bar{\mathbf{H}}}_k\!+\! \xi_k{\mathbf{I}}){\mathbf{w}}_{j} \!+\!  {\mathbf{z}}^{H}({\bar{\mathbf{H}}}_k\!+\! \xi_k{\mathbf{I}}){\mathbf{z}} \!+\!  {\mathbf{q}}^{H}({\bar{\mathbf{G}}}_k  \nonumber \\
&&\!\!\!\!\!\!\! + \tilde{\xi}_k{\mathbf{I}}){\mathbf{q}} +  \frac{{\delta_k^2}}{{\rho_k}}   \leq F_{{\bar{\mathbf{H}}}_{\xi_s,k},1}(\mathbf{w}_k,r_1,\tilde{\mathbf{w}}_k,\tilde{r}_1), \label{eq:zzy17a} \\
&&\!\!\!\!\!\!\!\!\!\!\!\!\!\!\!\!\! {{\sigma_l^2 \!+\!  {\mathbf{z}}^{H}({\hat{\mathbf{H}}}_l\!+\! \alpha_l{\mathbf{I}}  ){\mathbf{z}} \!+\! {\mathbf{w}}_k^{H}({\hat{\mathbf{H}}}_l\!+\! \alpha_l{\mathbf{I}}  ){\mathbf{w}}_k \!+\! {\mathbf{q}}^{H}({\hat{\mathbf{G}}}_l\!+\!  \tilde{\alpha}_l{\mathbf{I}}  ){\mathbf{q}}}}  \leq   \nonumber \\
&&\!\!\!\!\!\!\!\!\!\!\!\!\!\!\!\!\!\!\!  \sigma _l^2(\frac{2}{\tilde{r}_2}  - \frac{r_2}{\tilde{r}^2_2})\!+\!   F_{{\hat{\mathbf{H}}}_{\xi_e,l},0}({\mathbf{z}},r_2,{\tilde{{\mathbf{z}}}},\tilde{r}_2)\!+\!   F_{{\hat{\mathbf{G}}}_{\tilde{\xi}_e,l},0}({\mathbf{q}},r_2,{\tilde{{\mathbf{q}}}},\tilde{r}_2). \label{eq:zzy17b}
\end{eqnarray}
\end{subequations}

In order to approximate the EH constraint \eqref{eq:7b} and \eqref{eq:7c} to convex one,
we apply an SCA-based method. First, by using a loose approximation approach for \eqref{eq:7b} and \eqref{eq:7c}, we have
\begin{subequations}\label{eq:zzy188}
\begin{eqnarray}
&&\!\!\!\!\!\!\!\!\!\!  \eta_{c,k}({1 \!-\! \rho_k})(\sum\nolimits_{j=1}^{K}{\mathbf{w}}_{j}^H\big({\bar{\mathbf{H}}}_k\!+\! \Delta_k){\mathbf{w}}_{j} \!+\!  {\mathbf{z}}^{H}({\bar{\mathbf{H}}}_k\!+\! \Delta_k){\mathbf{z}}  \nonumber\\
&&\!\!\!\!\!\!\!\!\!\! + {\mathbf{q}}^{H} ({\bar{\mathbf{G}}}_k\!+\! \Phi_k)  {\mathbf{q}}+\sigma_k^{2}\big) \geq {\bar{E}_s}, \forall k, \label{eq:zzy188a}\\
&&\!\!\!\!\!\!\!\!\!\!
 {\mathbf{z}}^{H}({\hat{\mathbf{H}}}_l\!+\! \Upsilon_l  ){\mathbf{z}} \!+\! \sum\nolimits_{j=1}^{K}{\mathbf{w}}_k^{H}({\hat{\mathbf{H}}}_l\!+\! \Upsilon_l  ){\mathbf{w}}_k + {\mathbf{q}}^{H}({\hat{\mathbf{G}}}_l\!+\!  \Psi_l  ){\mathbf{q}} \nonumber\\
&&\!\!\!\!\!\!\!\!\!\!
  \geq {{\check{E}}}_e\!-\!  N_{E}\sigma_l^{2}, \forall l, \label{eq:zzy188b}
\end{eqnarray}
\end{subequations}
where ${{\check{E}}}_e = \frac{\bar{E}_e}{\eta_{e,l}}$.

Using loose approximation \cite{loose_approxi} to the LHS of \eqref{eq:zzy188a} and \eqref{eq:zzy188b} yields
\begin{subequations}\label{eq:zzy18}
\begin{eqnarray}
&&\!\!\!\!\!\!\!\!\!\!  \eta_{c,k}({1 \!-\! \rho_k})\big(\sum\nolimits_{j=1}^{K}{\mathbf{w}}_{j}^H({\bar{\mathbf{H}}}_k\!-\! \xi_k{\mathbf{I}}){\mathbf{w}}_{j} \!+\!  {\mathbf{z}}^{H}({\bar{\mathbf{H}}}_k\!-\! \xi_k{\mathbf{I}}){\mathbf{z}}  \nonumber\\
&&\!\!\!\!\!\!\!\!\!\! + {\mathbf{q}}^{H}({\bar{\mathbf{G}}}_k\!-\! \tilde{\xi}_k{\mathbf{I}}){\mathbf{q}}+\sigma_k^{2}\big) \geq {\bar{E}_s}, \forall k, \label{eq:zzy18a}\\
&&\!\!\!\!\!\!\!\!\!\!
 {\mathbf{z}}^{H}({\hat{\mathbf{H}}}_l\!-\! \alpha_l{\mathbf{I}}  ){\mathbf{z}} \!+\! \sum\nolimits_{j=1}^{K}{\mathbf{w}}_k^{H}({\hat{\mathbf{H}}}_l\!-\! \alpha_l{\mathbf{I}}  ){\mathbf{w}}_k  \nonumber\\
&&\!\!\!\!\!\!\!\!\!\! + {\mathbf{q}}^{H}({\hat{\mathbf{G}}}_l\!+\!  \tilde{\alpha}_l{\mathbf{I}}  ){\mathbf{q}}
  \geq {{\check{E}}}_e \!-\!  N_{E}\sigma_l^{2}, \forall l. \label{eq:zzy18b}
\end{eqnarray}
\end{subequations}

It is observed that ${\mathbf{w}}_{j}^H({\bar{\mathbf{H}}}_k\!-\! \xi_k{\mathbf{I}}){\mathbf{w}}_{j}$, ${\mathbf{z}}^{H}({\bar{\mathbf{H}}}_k\!-\! \xi_k{\mathbf{I}}){\mathbf{z}}$ and ${\mathbf{q}}^{H}({\bar{\mathbf{G}}}_k\!-\! \tilde{\xi}_k{\mathbf{I}}){\mathbf{q}}$ are the concave part of constraints \eqref{eq:zzy18a} and \eqref{eq:zzy18b}.  In order to make \eqref{eq:zzy18a} and \eqref{eq:zzy18b} more tractable, we employ the SCA technique for \eqref{eq:zzy18a} and \eqref{eq:zzy18b} to obtain convex approximations.

Firstly, we take ${\mathbf{z}}^{H}({\bar{\mathbf{H}}}_k\!-\! \xi_k{\mathbf{I}}){\mathbf{z}}$ as an example.  Let $\tilde{{\mathbf{z}}}$ be an initial feasible point.
We substitute  ${\mathbf{z}} \!=\!  {\tilde{\mathbf{z}}}\!+\! \Delta \mathbf{z}$ into ${\mathbf{z}}^{H}({\bar{\mathbf{H}}}_k\!-\! \xi_k{\mathbf{I}}){\mathbf{z}}$ as follows
 \begin{equation}\label{eq:zzy166}
\begin{split}
&{\mathbf{z}}^{H}({\bar{\mathbf{H}}}_k\!-\! \xi_k{\mathbf{I}}){\mathbf{z}} \\
=&({\tilde{\mathbf{z}}}\!+\! \Delta \mathbf{z})^{H}{\bar{\mathbf{H}}}_{\xi_s,k}({\tilde{\mathbf{z}}}\!+\! \Delta \mathbf{z}) \\
\geq & {\tilde{\mathbf{z}}}^{H}{\bar{\mathbf{H}}}_{\xi_s,k}{\tilde{\mathbf{z}}}+ 2\Re \{ {\tilde{\mathbf{z}}}^H{\bar{\mathbf{H}}}_{\xi_s,k}\Delta {{\mathbf{z}}}\},
\end{split}
 \end{equation}
where \eqref{eq:zzy166}  are derived by dropping the quadratic form $ {\Delta{\mathbf{z}}}^{H}{\bar{\mathbf{H}}}_{\xi_s,k}{\Delta{\mathbf{z}}}$.

Then, defining $\tilde{{\mathbf{w}}}_k$  and $\tilde{{\mathbf{q}}}$ as initial feasible point. Substituting ${\mathbf{w}}_k \!=\! {\tilde{\mathbf{w}}}_k\!+\!\Delta \mathbf{w}_k, \forall k,$ ${\mathbf{z}} \!=\!  {\tilde{\mathbf{z}}}\!+\! \Delta \mathbf{z}$, and ${\mathbf{q}} \!=\! {\tilde{\mathbf{q}}}\!+\! \Delta \mathbf{q}$  into the LHS of  \eqref{eq:zzy18a} and \eqref{eq:zzy18b}. Then, we can use the the similar method with \eqref{eq:zzy166} to achieve linear approximations
of the concave constraints \eqref{eq:zzy18a} and \eqref{eq:zzy18b}, respectively, as
\begin{subequations}\label{eq:zzy19}
\begin{eqnarray}
&& \eta_{c,k}a_k({1 \!-\! \rho_k}) \geq {\bar{E}_s}, \forall k, \label{eq:zzy19a}\\
&&\!\!\!\!\!\!\!\!\!\!\!\!\!\!  \sum\nolimits_{j=1}^{K} \left[ {\tilde{\mathbf{w}}}_{j}^H{\hat{\mathbf{H}}}_{\xi_e,l}{\tilde{\mathbf{w}}}_{j} \!+\! 2\Re \{ {\tilde{\mathbf{w}}}_{j}^H{\bar{\mathbf{H}}}_{\xi_e,l}\Delta {{\mathbf{w}}}_{j}\} \right] \!+\! 2\Re \{ {\tilde{\mathbf{z}}}^H{\bar{\mathbf{H}}}_{\xi_e,l}\Delta {{\mathbf{z}}}\}\nonumber \\
&&\!\!\!\!\!\!\!\!\!\!\!\!\!\!\!   \!+\!  {\tilde{\mathbf{z}}}^{H}{\bar{\mathbf{H}}}_{\xi_e,l}{\tilde{\mathbf{z}}} \!+\! {\tilde{\mathbf{q}}}^{H}{\hat{\mathbf{G}}}_{\tilde{\xi}_e,l}{\tilde{\mathbf{q}}}\!+\! 2\Re \{ {\tilde{\mathbf{q}}}^H{\hat{\mathbf{G}}}_{\tilde{\xi}_e,l}\Delta {{\mathbf{q}}}\}
\geq {{\check{E}}}_e \!-\! N_{E}\sigma_l^{2}, \label{eq:zzy19b}
\end{eqnarray}
\end{subequations}
where $a_k \!=\! \sum\nolimits_{j=1}^{K} \left[ {\tilde{\mathbf{w}}}_{j}^H{\bar{\mathbf{H}}}_{\xi_s,k}{\tilde{\mathbf{w}}}_{j} \!+\! 2\Re \{ {\tilde{\mathbf{w}}}_{j}^H{\bar{\mathbf{H}}}_{\xi_s,k}\Delta {{\mathbf{w}}}_{j}\} \right]+ {\tilde{\mathbf{z}}}^{H}{\bar{\mathbf{H}}}_{\xi_s,k}{\tilde{\mathbf{z}}} +2\Re \{ {\tilde{\mathbf{z}}}^H{\bar{\mathbf{H}}}_{\xi_s,k}\Delta {{\mathbf{z}}}\} +{\tilde{\mathbf{q}}}^{H}{\bar{\mathbf{G}}}_{\tilde{\xi}_s,k}{\tilde{\mathbf{q}}} +2\Re \{ {\tilde{\mathbf{q}}}^H{\bar{\mathbf{G}}}_{\tilde{\xi}_s,k}\Delta {{\mathbf{q}}}\}+ \sigma_k^{2}$ and ${\bar{\mathbf{G}}}_{\tilde{\xi}_s,k}={\bar{\mathbf{G}}}_k\!+\! \tilde{\xi}_k{\mathbf{I}}$.
In addition, it is noted that \eqref{eq:zzy19a} is still non-convex in its current form since it involve coupled $a_k$'s and $1 \!-\! \rho_k$'s. The first-order Taylor expansion of $\sqrt{\bar{E}_s}$ is
$e_s = \sqrt{{\tilde{E}_s}} +0.5{{\tilde{E}_s}}^{-\frac{1}{2}}({{\bar{E}_s}} -{\tilde{E}_s} )$.
Thus, \eqref{eq:zzy19a} can be rewritten to a convex second-order
cones (SOC) constraint as
\begin{equation}\label{zhu321}
 \left\|\left[ 2{{{e_s}} \mathord{\left/
 {\vphantom {{{e_s}} {\sqrt {{\eta _{c,k}}} }}} \right.
 \kern-\nulldelimiterspace} {\sqrt {{\eta _{c,k}}} }}, a_k+\rho_k-1\right]\right\|\leq  a_k-\rho_k+1.
\end{equation}

In addition, \eqref{eq:6c} can be reformulated to two SOC forms as
\begin{subequations}\label{zhengyu}
\begin{eqnarray}
&&\!\! \left\|\left[ \mathbf{w}_1^T, ..., \mathbf{w}_K^T, \mathbf{z}^T\right]\right\|\leq \sqrt{P_T},   \label{eq:zhengyua} \\
&& ~~~~  \left\| \mathbf{q}^T \right\|\leq \sqrt{P_J}. \label{eq:zhengyub}
\end{eqnarray}
\end{subequations}

Eventually, problem \eqref{eq:6} is converted into the following convex second order cone
programming (SOCP) problem as
\begin{equation}\label{zhu32}
 \begin{split}
& ~~~~ ~~~~  \max_{\rho_k,{\kern 1pt}\{{\mathbf{w}}_{k}\},{\kern 1pt}{\mathbf{z}},{\kern 1pt}{\mathbf{q}}, {\kern 1pt}{\tilde{E}_s},{\kern 1pt}\bar{E}_{e},{\kern 1pt}  {\kern 1pt}r_1, {\kern 1pt}r_2}  ~~~~ ~~  \tau \bar{E}_{s}  +   (1-\tau) \bar{E}_{e} ~~~  \\
& \mbox{s.t.}~ \eqref{eq:11a}, {\kern 1pt}\eqref{eq:zzy17a}, {\kern 1pt}\eqref{eq:zzy17b}, {\kern 1pt} \eqref{eq:zzy19b}, {\kern 1pt}\eqref{zhu321}, {\kern 1pt}\eqref{eq:zhengyua}, {\kern 1pt}\eqref{eq:zhengyub}, {\kern 1pt}0 < \rho_k \leq 1.
\end{split}
\end{equation}
Given $\{\tilde{{\mathbf{w}}}_{k}\}$, $\tilde{{\mathbf{q}}}$, $\tilde{{\mathbf{z}}}$, $\tilde{r}_1$,  $\tilde{r}_2$, and ${\tilde{E}_s}$ problem \eqref{zhu32} is convex and can be solved by employing  convex optimization software tools such as CVX \cite{27 CVX}. Based on the SPCA method, an approximation
with the current optimal solution can be updated iteratively,
which implies that \eqref{eq:6} is optimally solved.

%
%
%
%
%

\section{Simulation Results}
In this section, we provide the simulation results to validate the performance of our proposed schemes. We set that $K = 2$, $L = 2$, $M = 1$,  $N_T = 4$,  $N_J = 4$, and  $N_E = 2$.
We assume the estimated channel ${\overline {\mathbf{h}} _k}$, ${\overline {{\mathbf{H}}} _l}$,${\overline {\mathbf{g}} _k}$, and ${\overline {{\mathbf{G}}} _l}$ are respectively modelled as ${\overline {\mathbf{h}} _k} {\rm{ = }}H({d_k}){\overline {\mathbf{h}} _I}$, ${\overline {{\mathbf{{H}}}} _l} = H({d_l} ){\overline {{\mathbf{{H}}}} _I}$, ${\overline {\mathbf{g}} _k}=H{\kern 1pt} ({f_k}{\kern 1pt} ){\overline {\mathbf{g}} _I}$, and ${\overline {{\mathbf{G}}} _l} = H{\kern 1pt} ({f_l}{\kern 1pt} ){\overline {{\mathbf{G}}} _I}$, where ${\overline {\mathbf{h}} _I} \sim \mathcal{CN}(0,{\mathbf{I}})$, ${\overline {{\mathbf{H}}} _I} \sim \mathcal{CN}(0,{\mathbf{I}})$, ${\overline {\mathbf{g}} _k} = \sim \mathcal{CN}(0,{\mathbf{I}})$, and ${\overline {{\mathbf{G}}} _l} = \sim \mathcal{CN}(0,{\mathbf{I}})$, $H({d_k}) = \frac{c}{{4\pi {f_c}}}{(\frac{1}{{{d_k}}})^{\frac{\kappa }{2}}}$.  We define ${d_k} = 100 $  m  and ${f_k} = 100 $  m  meters as the distance between the transmitter as well as the CJ and all the  CR, and ${d_l}{\kern 1pt}  = 9 $ m and ${f_l}{\kern 1pt}  = 9$ m meters as the distance between the transmitter as well as the CJ and all the ER, unless otherwise specified. 
Moreover, $c = 3 \times {10^8}{\rm{m}}{{\rm{s}}^{ - 1}}$ is the speed of light; ${f_c} = 900$ MHz  is the carrier frequency; and $\kappa = 2.7$ is the path loss exponent.  In addition, the noise power at the CR is set to be $ \sigma_k^{2} = -90 $ dBm and $ \delta_k^2 = -50$ dBm. Also the noise power of all the ERs is $ \sigma_{k}^{2} = -90$ dBm, $\forall k$. Also we set  the channel error bound for the deterministic model as $  \varepsilon_{s} = \varepsilon_k = \tilde{\varepsilon}_k, \forall k$  and $ \varepsilon_{e} = \theta_l = \tilde{\theta}_l , \forall k$.  The EH efficiency coefficients are set to $\eta_{c,k} = \eta_{e,l} =$ 0.3 and  the priority parameter $\tau$ is 0.5.

In our simulations, we compare the following transmit designs: the perfect CSI case, the proposed SOCP-SPCA algorithm,  the no-CJ scheme which is the robust design w/o CJ by setting ${{\mathbf{Q}}}= {\mathbf{0}}$ \cite{TIFS_zhu},  the no-AN scheme which means the robust design w/o CJ by setting ${{\mathbf{Z}}} = {\mathbf{0}}$ \cite{23 Qin_Cooperative_jamming aided robust secure},   and the non-robust method which is a scheme that assumes no uncertainty in the CSI.

\begin{figure}[!htbp]
\centering
\includegraphics[scale = 0.4]{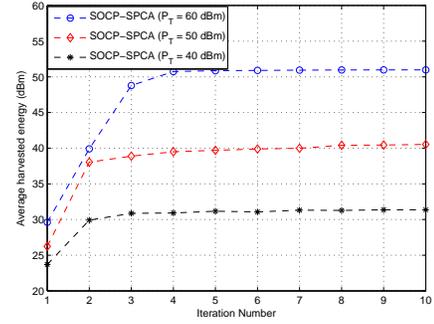}
\caption{Average harvested power versus iteration numbers}
\label{fig:power_VS_iternumber_161129}
\end{figure}

Fig. 1 illustrates the convergence of the SOCP-SPCA method with respect to iteration numbers for $ P_T = 40$ dBm,  $ P_J = 40$ dBm, $\bar{R}_s= 0.5$ bps/Hz, and $\varepsilon =$ 0.01, respectively. It is easily seen that convergence of the SOCP-SPCA method is achieved for all cases within just 5 iterations.

\begin{figure}[!htbp]
\centering
\includegraphics[scale = 0.4]{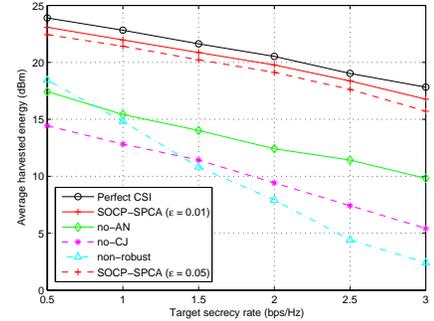}
\caption{Average harvested power versus target secrecy rate}
\label{fig:Har_power_VS_rate_170901}
\end{figure}

Fig. 2 shows the average harvested power in terms of different target secrecy rates with $ P_T = 30$ dBm and $ P_J = 30$ dBm, respectively. It is observed that the harvested power of all schemes decline with the increase of the secrecy rate target. Also, the performance gain of the scheme with $\epsilon = 0.01$ over the scheme with $\epsilon = 0.05$ is 0.8 dB at all the target secrecy rate region.
Compared with the no-AN scheme and the no-CJ scheme, the harvested power of the SOCP-SPCA algorithm are $6$ dB and $9$ dB higher. Moreover, we can check that the SOCP-SPCA algorithm perform better than the non-robust scheme, and the performance gap increases as the target secrecy rate becomes large. 

\begin{figure}[!htbp]
\centering
\includegraphics[scale = 0.4]{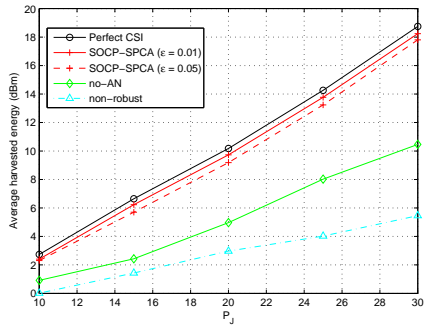}
\caption{Average harvested power versus the power budget at the CJ }
\label{fig:Har_power_VS_PJpower_170901}
\end{figure}

Fig. 3 depicts the average harvested power versus the power budget at the CJ with $ P_T = 10$ dBm and $\bar{R}_s= 0.5$ bps/Hz, respectively.  It is easily observed that the achieved
harvested power increases with $P_J$, and the curves of the perfect CSI case and the SOCP-SPCA algorithm increase with the same slope.
Moreover,  we can check that as $ P_J $ increases, the performance gap between the proposed algorithms and the no-AN scheme becomes larger and the performance loss of the non-robust scheme grows.


\section{Conclusion}
In this paper, we have studied the robust secure beamforming design for a
MU-MIMO SWIPT secrecy system with the PS scheme by incorporating
the norm-bounded channel uncertainties.  We aim to maximize the minimum of harvested energy by jointly optimizing the AN-aided beamforming, the AN vector, the CJ vector and the PS ratio design.
To solve the non-convex problem, we use the SPCA method,  loose
approximation and SCA-based method to reformulate the original problem as an convex SOCP problem.
Also, an SPCA-based iterative algorithm is also addressed.
Finally, simulation results have been provided to validate the performance of our proposed algorithm.  In addition, the proposed robust design methods outperforms the
non-robust schemes.

%
%

\ifCLASSOPTIONcaptionsoff
  \newpage
\fi



%
%
%

\end{document}